%% file: main.tex
\documentclass[11pt]{article}
\usepackage{UF_FRED_paper_style}
\usepackage{hyperref}
\hypersetup{breaklinks=true}
\usepackage{breakurl}

\usepackage{lipsum}  

\usepackage{xcolor}
\usepackage{markdown} 
\usepackage{booktabs}
\usepackage{multirow}
\usepackage{graphicx}
\usepackage{authblk}
\usepackage{enumitem}

\newlist{checklist}{itemize}{1}
\setlist[checklist]{label=$\square$, topsep=0pt, partopsep=0pt, parsep=0pt, itemsep=0pt}

\title{Web Scraping for Research: Legal, Ethical, Institutional, and Scientific Considerations
}

\author[1,2]{Megan A. Brown\thanks{Corresponding Author: mgnbrown[at]umich.edu}}
\author[3]{Andrew Gruen}
\author[4]{Gabe Maldoff}
\author[2]{Solomon Messing}
\author[2,5]{Zeve Sanderson}
\author[6]{Michael Zimmer}
\affil[1]{School of Information, University of Michigan}
\affil[2]{Center for Social Media and Politics, New York University}
\affil[3]{Mozilla Foundation \textcolor{white}{,}}
\affil[4]{University of Maine School of Law}
\affil[5]{Vrije Universiteit Amsterdam}
\affil[6]{Department of Computer Science, Marquette University}

\date{\today}

\begin{document}

\pagenumbering{arabic}
{\setstretch{.8}
\maketitle
\begin{abstract}

Scientists across disciplines often use data from the internet to conduct research, generating valuable insights about human behavior. However, as generative AI relying on massive text corpora becomes increasingly valuable, platforms have greatly restricted access to data through official channels. As a result, researchers will likely engage in more web scraping to collect data, introducing new challenges and concerns for researchers. This paper proposes a comprehensive framework for web scraping in social science research for U.S.-based researchers, examining the legal, ethical, institutional, and scientific factors that we recommend researchers consider when scraping the web. We present an overview of the current regulatory environment impacting when and how researchers can access, collect, store, and share data via scraping. We then provide researchers with recommendations to conduct scraping in a scientifically legitimate and ethical manner. We aim to equip researchers with the relevant information to mitigate risks and maximize the impact of their research amidst this evolving data access landscape.

\noindent

\end{abstract}
}
\vspace{1cm}
\textbf{Legal Disclaimer}
This paper is intended for informational and academic purposes only. The statements made regarding the law, ethics, and logistics of web scraping are based on general understanding and research as of the publication date. This document does not constitute legal advice, nor does it create an attorney-client relationship. Laws and regulations surrounding web scraping may vary by jurisdiction and are subject to change. Readers are encouraged to consult qualified legal professionals to obtain advice specific to their individual circumstances before engaging in any scraping activities.

\newpage
\section{Introduction}

In 2023, Twitter announced it would suspend no-cost academic research access to its API, with a new pricing structure would require researchers to pay roughly \$500,000 per year for comparable access \citep{twitterdev}.\footnote{Researcher API access previously provided 10 million tweets per month. The Basic access tier announced on March 29, 2023 provides 10 thousand tweets for \$100 per month; the Pro tier provides one million tweets for \$5,000 per month; and the small enterprise access provides 50 million tweets for \$42,000 per month, making Twitter cost-prohibitive to most researchers.} 
In April 2023, further restrictions were imposed when Twitter deactivated developer accounts set up under its former policy, stalling numerous research projects and public initiatives that relied on Twitter data \citep{citrletter}.
In August 2024, Meta shut down CrowdTangle less than three months before the 2024 U.S. elections and in the midst of elections in many other countries around the world \citep{ctdead}.
This tool previously enabled researchers to access public data via an Application Programming Interface (API) and dashboard, offering the ability to download and analyze this data independently.  
Consequently, social media researchers and those who built public goods on top of those data were left with few options for data collection through official channels. 

At the same time, the regulatory environment governing data scraping remains unclear. 
There exists a patchwork of state and federal regulations regarding both access and storage of internet data, particularly when that data involves personally identifiable information. 
Additionally, new data access laws being considered in the U.S. and going into effect in the E.U. expand avenues for researchers to scrape data. 

Given this backdrop, we expect to see an increase in research relying on scraping data from the web---including by researchers new to the practice---which creates a need for guidelines outlining the opportunities and risks associated with scraping.
To guide researchers who may be taking on scraping projects for the first time, we propose a framework for understanding the multifaceted landscape of scraping for social science research, examining the legal, ethical, institutional, and scientific considerations of data access via scraping. 
Using this framework, we identify critical questions researchers must consider to ensure that research is scientifically rigorous and ethical while minimizing legal risks when using scraping as a data collection tool. 
We provide guidance on key questions that researchers will face with these data collection methods, recognizing that some of the challenges are immediately resolvable while others will likely require multi-stakeholder coordination and consensus building that leads to the creation of new norms, practices, and regulatory frameworks. 
We also provide a check list for researchers to think through the considerations we outline. 
Taken together, the information and recommendations in this paper can advise researchers on how to navigate the complexities of scraping while maximizing the potential for ethical, high-impact, scientifically rigorous research. 
    
\section{What is scraping?}

We define scraping to be automated data collection via the internet that captures data designed to be used and/or rendered on a web page or app. 
The study of social media is rich with examples of work based on scraped datasets (e.g. see \cite{faddoul2020longitudinal, aslett2024online, bisbee2022election, robertson2018auditing, boeker2022empirical, baumgartner2020pushshift,lange2022spreading, lan2021covid,fathoni2022mapping}, but the practice goes back to nearly the beginning of the Web. 
    
The ``World Wide Web Wanderer" created in 1993 at MIT may be the earliest incidence of scraping.  
Researchers interested in the growth of the Web built a tool that ``spider-walked'' the web, following link to link to the end of a chain and recording information about the number of unique pages discovered \citep{gray1995webgrowth}.  
This approach is employed to this day in the form of ``crawlers''---tools used most often by search engines to collect the contents of web pages, index them, and then rank them by search terms.
  
In the early days of its development, Microsoft's Bing scraped search results from Google to improve its product. 
When Google caught Bing, Microsoft vice president Harry Shum admitted to doing so (along with monitoring clicks on Google in its browser), arguing that everyone collectively provides the data for search, and pointing out that across technology companies, ``Everyone does this..." \citep{singel_google_2011}. 
    
Today, norms around scraping in the technology industry are broad and flexible. 
Google, Microsoft, Meta, OpenAI, and other tech companies large and small regularly scrape and index content from across the web on a continual basis, for profit, and without explicit permission; in the case of machine learning model builders, this practice is increasingly contentious  \citep{grynbaum_times_2023}.\footnote{Most large companies do not index websites if the \texttt{robots.txt} file opts out.  
\texttt{Robots.txt} was codified in 1994 by Martijn Koster, and is only now going through a standards codification process via the Internet Engineering Task Force's Request for Comments process \citep{rfc9309}.} 
Sometimes companies profit by way of public-facing free services, which they monetize via advertising, as is the case with most search engines (e.g. Google or Bing). 
However, scraped data can be used more directly for profit: Companies also scrape to develop structured data they can sell---for everything from checking academic publications for plagiarism \citep{turnitin} to collecting social media photos to create facial recognition systems for police \citep{clearviewai} to building chatbots for which customers may directly pay \citep{openai}. 
Moreover, some companies simply sell the scraped data themselves.\footnote{Some of the highest-profile legal cases involving scraping have focused on simply selling scraped data. For example, \textit{hiQ Labs, Inc v. LinkedIn Corporation}, one of the only federal appellate decisions to address scraping, involved access hiQ's access to LinkedIn data to create a "people analytics" product, which it marketed and sold to business customers \citep{hiQLabs2022}.} 
    
The practice is not limited to web crawling for search engines, AI model builders, or enterprise data brokers.  Scraping has become a primary method of data collection in many industries for both public interest work and for-profit work. 
Journalists regularly scrape to investigate and tell stories, arguing it is central to a functioning democracy \citep{markup2020democracy}. 
In other cases, activists have scraped court records to make analysis and oversight of powerful entities, such as law enforcement, easier \citep{pdap}. 
Despite its broad application, the ethical and legal frameworks governing scraping remain ambiguous and often contentious, underscoring the need for clearer guidelines for researchers.

\subsection{How do researchers scrape data?}
Researchers can scrape data using several mechanisms \citep{davidson2023platform,ohme2024digital}. 
Such mechanisms include traditional scraping, which involves the \textit{creation of structure} from inconsistently or loosely structured data from a website, often by parsing HTML.  
Researchers can also scrape data via undocumented APIs (Application Programming Interfaces), where researchers use the APIs that serve web content to end users, which are typically unofficial, undocumented, and not permitted for use by third party developers. 
Lastly, researchers can enroll participants or ``citizen scientists'' in the use of a browser plugin, which collects data from a participants' web activity and relays it to the research team. 
Importantly, these three access methods are distinct from the official APIs provided by websites (e.g. the Meta Content Library or Twitter's Developer API), which platforms explicitly make available for developers to interact with platform data under conditions set by the platforms. 
For researchers interested in more detail regarding these methods, we provide details in Appendix~\ref{SI:accessmethods}. 
Each of these mechanisms for scraping---traditional scraping, undocumented API scraping, and browser plugin scraping---have different legal, scientific, ethical, and institutional factors that researchers must consider when conducting their projects, which we describe below.

\section{Research Considerations}
In this section, we provide an overview of the legal, ethical, and scientific considerations around scraping. We first define the legal considerations for researchers taking on a scraping project, focusing on data access laws that impact researchers in the United States, including U.S. criminal and civil statues as well as laws outside the U.S. that researchers should consider.
We then discuss the ethics of research using scraped data and how researchers can assess risks to subjects in the data. 
Given that many researchers work within university contexts, we provide an overview of institutional resources and constraints that researchers must consider when scraping data. 
Lastly, we discuss scientific factors regarding the validity of scraped data that should inform research design. 

\subsection{Legal Considerations} \label{sect:legal}
Scraping – like the ``law of the horse\footnote{When, in 1996, Judge Frank Easterbrook was asked to present on “The Law of Cyberspace,” he questioned whether it was appropriate to talk of a “cyberspace” or “Internet” law at all and he likened such a concept to a “law of the horse" \citep{Easterbrook1996}. Like Internet law, scraping is situational and implicates disparate areas of law, but common themes and trends emerge that can guide researchers through the maze of legal complexity.}'' – implicates several overlapping areas of law and regulation. First, behavior on the Internet is often mediated by contractual terms enforced by various private actors that create the content and virtual spaces that form the Internet. Second, scraping can intersect with various statutes and common laws that regulate intellectual property, trespassing, and computer hacking. Third, where scraping results in the collection of personal information, scraping can intersect with privacy and data protection laws that may dictate notice and choice requirements as well as technical and organizational protections that could interfere with, or at least complicate, some areas of research. 

All of these areas of law continue to evolve with the growth of data-driven business models and increasing recognition of the relationship between data and economic power (e.g. see \citet{Fung_2023}).
To add complexity, the global nature of the internet means that different countries’ laws may apply to scraping depending on various factors, such as the location of those doing the scraping, the location of the data source, and the location of any individuals whose data are scraped.

\subsubsection{Contractual restrictions}
Social media companies and other website operators rely on agreements with their users – often titled ``terms of service,'' ``terms of use,'' ``terms and conditions,'' and ``end user license agreements'' (collectively, ``user contracts'') – to protect their legal rights, including to police intellectual property interests, set acceptable guidelines for user behavior, and define dispute resolution procedures. User contracts may seek to regulate and/or prohibit scraping and other forms of automated data collection or analysis necessary for research purposes \citep{Sobel2021}. At the same time, as this section explains, courts have increasingly limited the power of platforms and websites to unilaterally restrict scraping through take-it-or-leave-it user contracts. 
            
\paragraph{Restrictions on scraping}
Most social media companies, and many other websites, prohibit scraping and other automated data collection without the company’s permission. Whether framed as a restriction on the automated \textit{collection} of data from the company’s services, or as a prohibition on automated \textit{access} to the services, these terms are designed to limit scraping by restricting the \textit{means} by which the services or relevant data are accessed. While user contracts may restrict unauthorized data collection and other activities relevant to research in a variety of ways, provisions blocking scraping specifically tend to distinguish ``automated'' from non-automated uses of the services. 

User contracts may also implicitly authorize some forms of scraping without requiring a researcher to seek the company’s express permission. Most websites, including social media services that prohibit scraping in their user contracts, permit ``crawling" – automated scanning that allows search engines to link to and index a website. These websites may authorize such data collection via a \texttt{robots.txt} file that is accessible to search engines’ indexing bots. Some companies’ \texttt{robots.txt} files permit only specifically named third parties to crawl their services, while others allow crawling by any third party while strictly limiting which elements of the service the third party may crawl. The line between scraping and crawling is not clearly defined.  

\paragraph{Challenging restrictions on scraping}
The fact that a contract prohibits scraping is not the end of the inquiry. While contract law varies by state, for a website operator to assert there has been a breach of its user contract, it generally needs to show: (1) a valid agreement, (2) a breach of the terms of the agreement, and (3) resulting damages to website operator \citep{MetaPlatforms2022}. In recent years, courts have begun to question the circumstances under which a user contract forms a valid agreement between a website operator and a party that scrapes the website. This analysis depends on how the scraping party ``accepted'' the agreement and the nature of the scraping that took place \citep{XCorpVBrightData2023}.

User contracts generally state that users agree to the contract when the user creates an account, logs in to the service, or simply uses the website \citep[p.~1256]{MetaPlatforms2022}. 
``Browsewrap'' agreements imply user consent to terms simply by using a service, while ``clickwrap" agreements require explicit user action, such as clicking a button to accept. A hybrid, ``sign-in-wrap" agreements, obtain consent when a user signs up for or logs into a service.

Courts in a number of jurisdictions are beginning to scrutinize browsewrap agreements more closely, particularly where such agreements impose one-sided provisions that courts consider unfair or unexpected. For example, in the context of scraping, the Court of Appeals for the Ninth Circuit has explained, ``[u]nless the website operator can show that a consumer has actual knowledge of the agreement, an enforceable [browsewrap] contract will be found based on an inquiry notice theory only if (1) the website provides reasonably conspicuous notice of the terms to which the consumer will be bound; and (2) the consumer takes some action, such as clicking a button or checking a box, that unambiguously manifests his or her assent to those terms'' \citep{Berman2022}. 

Even where the scraping party has taken an affirmative action to agree to the user contract – including in the case of clickwrap and sign-in-wrap agreements  – some courts have refused to enforce anti-scraping provisions where the scraping was unrelated to the user's agreement to the user contract. For example, one California court held that a platform's user contract did not restrict scraping by a party that had agreed to the platform's user contract because the scraping took place through tools that could access the platform when the party was not logged in \citep{MetaPlatforms2024}. 
In other words, even though the court acknowledged that there was a valid user contract that restricted scraping, the court determined that the contract did not control behavior that occurred outside of the scraping party's use of the service as a ``user'' because the scraped data would have been visible even to non-users who had not accepted the platform's user agreement.

Moreover, even where a valid contract does bind the relevant scraping, website operators may face additional challenges when seeking to enforce the user contract in the context of research.  A key hurdle for website operators asserting breach of contract claims is to demonstrate that the operator was \textit{harmed} by the breach. Platforms typically assert harm in commercial scraping cases by alleging that scraping results in unfair competition. However, competition considerations may be less persuasive outside the commercial context.  In at least one case involving scraping by a non-profit entity, a court found that the non-profit’s scraping of publicly available information for advocacy purposes did not harm the platform because the relevant information was already accessible to the public \citep{XCorp2024}. 
Courts may also be willing to accept public interest-related defenses to scraping for research purposes. For example, while courts have generally not entertained freedom of expression defenses in cases involving website scraping for commercial purposes, in the research context, free-expression arguments have not been tested and may have more currency. 

Researchers should exercise caution before concluding that they are not bound by browsewrap agreements that restrict scraping. For starters, in the Berman v Freedom Financial Network, LLC decision, the court noted that if ``the website operator can show that a consumer has actual knowledge of the agreement'' the terms prohibiting scraping may be enforceable \citep[p.~856]{Berman2022}. 
What's more, cases where courts have set aside anti-scraping provisions have typically involved fact patterns where the defendant scraped only publicly-available information or information that the relevant user was already permitted to access through ordinary use of the service. 
Courts may be less willing to permit scraping that involves content protected by technical safeguards and/or accessible only to certain classes of users.\footnote{In this context, researchers' deploying other users to assist in scraping -- such as through the installation of browser extensions that can scrape within their user accounts -- could lead to breach of contract claims, at least with regard to those users who have accepted the platform's user agreement.}

\subsubsection{Statutory restrictions on scraping}
Aside from contractual restrictions, scraping can also raise statutory considerations. For example, the Computer Fraud and Abuse Act (``the CFAA'') has been used by companies and website operators attempting to shut down scraping. The CFAA is a federal law originally drafted to prohibit certain forms of hacking. It states: ``[w]hoever...intentionally accesses a computer without authorization or exceeds authorized access, and thereby obtains...information from any protected computer...shall be punished by fine or imprisonment'' \citep{USC1030a2C}. 
While these laws can technically impose criminal penalties on scraping, ``the real area of growth has been with the CFAA’s civil provisions'' \citep{Sellars2018}. 

Website operators may bring CFAA claims to attempt to shut down data scraping activity by arguing that such scraping involves access to the website without authorization. For a CFAA claim to succeed, a website operator must show: (1) unauthorized access to the platform or exceeding authorized access and (2) loss of at least \$5,000 as a result of the violation \citep{FacebookPower2016}. 
In the research context, the United States District Court for the District of Columbia has ruled that a researcher's violation of a website's user contract alone is insufficient to generate criminal liability if the researcher did not breach a website's technical controls in order to gain ``unauthorized access'' \citep{Sandvig2020}.

In evaluating whether there has been ``unauthorized access,'' courts look at the technical structure for access to content on the website \citep{hiQLabs2022}. 
For example, where scraping occurs on a website that is available to the general public – without the need for an account or any other authorization – courts have found that the CFAA does not apply \citep[p.~1261--62]{MetaPlatforms2022}. 
For password-protected web pages, by contrast, the CFAA could apply to access without appropriate authorization \citep[p.~1199-1200]{hiQLabs2022}, 
but courts have found that access to such web pages through a valid user account is sufficient to demonstrate that access was ``authorized'' and not regulated by the CFAA \citep[p.~1067]{FacebookPower2016}.\footnote{This case held that Power Ventures did not  access Facebook without authorization when users initially gave Power permission to access Facebook data through their accounts. Rather, Power Ventures only violated CFAA when they continued to access Facebook data after Facebook sent a cease and desist and blocked Power Ventures IP addresses from accessing Facebook's servers.} 
At the same time, accessing content through a valid user account is not a permanent shield from CFAA liability because website operators can rescind such permission (including by revoking user accounts, by setting IP address blockers, or by sending cease-and-desist letters), after which continued access could constitute a CFAA violation \citep{FacebookPower2016}.

\subsubsection{Privacy and data protection laws}
Where scraping involves personal information, researchers may be directly regulated by privacy and data protection laws. U.S. privacy laws provide significant exceptions for ``publicly available information,'' for non-profit organizations, and for certain forms of research, all of which limit the application of U.S. privacy laws to most academic research. However, outside the U.S., many jurisdictions, including the European Union, apply broad protections to personal information that do not exclude researchers or publicly available information. These laws can apply extraterritorially and therefore may be relevant to U.S.-based researchers. 
            
\paragraph{U.S. privacy laws} 
There is no single federal privacy law that applies across all sectors, but rather a patchwork of federal and state requirements that govern the collection and processing of personal information. At the federal level, specific laws apply to specific sectors (e.g. the Health Insurance Portability and Accountability Act (``HIPAA'') applies to covered entities in the healthcare sector). Additionally, the Federal Trade Commission (FTC) has general authority to police ``unfair and deceptive acts or practices,'' which the FTC has wielded to shape norms affecting how companies collect, use, and share personal information \citep{FTC2024}. 
However, this authority has not been invoked to regulate academic research \citep{Levine2021}. 

At the state level, many states have passed consumer privacy laws, such as the California Consumer Privacy Act (CCPA). As of this writing, seventeen states have passed consumer privacy laws modeled to varying degrees on the CCPA. Most, but not all, of these state privacy laws include important exemptions for non-profit entities. Additionally, state laws vary with respect to the definition of ``publicly available'' data which the laws govern. We elaborate on these dimensions in Appendix~\ref{SI:usprivacylaws}. 

To the extent that these U.S. laws apply to scraping for research purposes, researchers generally can comply by maintaining public-facing privacy policies that describe their collection, use, and disclosure of personal information, and by permitting individuals to gain access to a copy of their personal information on request. Researchers may also be subject to other requirements from their respective institutions and research teams regarding how data is collected, analyzed, and stored, such as to implement appropriate security to protect personal information and to apply ``data minimization'' standards by collecting and storing only information that is reasonably necessary for the intended research purposes. These laws also provide individuals with the right to request that researchers delete personal information, but in many cases a researcher may have grounds to override the request under various exceptions

\paragraph{Data protection laws in other countries}
Privacy and data protection laws outside the U.S. may be implicated where research impacts individuals in other countries or where research teams have a presence in those countries. For example, the Personal Information Protection Law in China, the General Data Protection Regulation in the EU, and the General Data Protection Law in Brazil all regulate the use of data on individuals within their jurisdictions (though frequently with provisions accounting for academic research). Because the EU General Data Protection Regulation (``GDPR'') has become a model for data protection law around the world (and is known as one of the world’s strictest data protection regimes), we use the GDPR as an example of how such laws may impact data scraping for research.

Researchers may be subject to requirements under GDPR, even if the research team has no link to the EU, if they are ``monitoring the behavior" of individuals in the EU, such as by tracking or evaluating any individual's activities in the EU over time. Key provisions in the GDPR require entities to establish a ``legal basis'' for processing personal information, to provide a privacy notice to individuals whose data the business processes, and to offer those individuals certain privacy rights regarding the data, among other internal accountability and documentation requirements. These requirements can pose challenges to researchers that are attempting to scrape data. However, GDPR provides important exemptions and relaxed requirements for researchers. 

First, researchers can demonstrate that processing personal information serves lawful and legitimate purposes without creating undue risks for individuals, creating a ``legitimate interests legal basis'' for processing said data. Often, research meeting ethical and methodological standards will satisfy this legal basis, but the researcher bears the burden of documenting that the benefits outweigh the risks to individuals \citep{EDPS2020}. Researchers can mitigate risk by appropriately safeguarding the data -- including through technical security controls and through privacy-enhancing techniques like pseudonymization -- and minimizing the data collected to only what is strictly necessary \citep{EDPBChatGPT2024}. Researchers may also be exempt from requirements mandating privacy notices due to the disproportionate effort that would be required to do so \citep{GDPR_Art145b}. Lastly, researchers may be exempt from data subject rights to the extent that those rights limit legitimate research activities (e.g. where individuals opt-out and skew research results) \citep{EDPS2020}. However, like other exemptions, the law puts the onus on researchers to demonstrate how the proposed research design appropriately addresses privacy risks for the data subjects including appropriate technical and organizational measures \citep{GDPR_Art891}. We provide further detail on these requirements in Appendix~\ref{SI:GDPR}.

\subsubsection{Data access laws}
Data access laws may provide a legal basis for scraping publicly available data. 

The E.U. Digital Services Act (the ``DSA''), which passed in 2022, introduced a new regime designed to facilitate access to data held by certain ``very large online platforms'' and ``very large online search engines'' to study ``conduct research that contributes to the detection, identification and understanding of systemic risks in the Union." 

Under the DSA, designated very large online platforms and search engines must allow researchers to access data that is ``publicly accessible'' in a platform or search engine's ``online interface''--including, potentially, where access occurs via scraping. Access to such publicly available data is not limited to researchers affiliated with a qualifying research organization and would extend to researchers affiliated with nonprofit organizations \citep{DSA_Art4012}.

In addition, the DSA permits a ``vetted researcher'' to access data held by platforms that is \textit{not} publicly available, provided the researcher meets certain conditions\citep{DSA_Art404}. 
Specifically, under Article 40 of the DSA, a researcher must demonstrate to a national regulator—the ``Digital Services Coordinator''—that:

\begin{itemize}
    \item the researcher is affiliated with a qualifying research organization;
    \item the researcher is independent from commercial interests and has disclosed all funding sources;
    \item the researcher is capable of adequately protecting the privacy and security of the relevant data;
    \item access to the data is necessary and proportionate to the intended research purpose; and 
    \item the research will be made publicly available free of charge \citep{DSA_Art408}.
\end{itemize}

The DSA's data access regime is not expressly limited to researchers in the E.U. Rather, such provisions may apply to researchers anywhere — including in the U.S. — provided that the purpose of the research is to study ``systemic risks in the Union.''\footnote{Note, however, that it is likely E.U.-based researchers will be prioritized by various regulatory authorities who must grant vetted researcher status---in particular, the Digital Services Coordinators.} For research programs granting access to public data or scraping data with the permission of the platforms, some platforms currently make the data available to researchers in all jurisdictions, not just the E.U. Currently, there is no standard by which platforms include or exclude U.S.-based researchers in these programs.

The U.S., by contrast, does not currently have similar laws that grant researchers rights to access data held by private parties. Although U.S. privacy laws contain important exceptions that facilitate the collection and use of personal information in the context of academic research, such laws do not compel private parties to share data with researchers. However, policymakers have begun to propose laws designed to broaden research data access. For example, had it passed, a 2023 bill titled ``The Platform Accountability and Transparency Act (PATA),'' would have created a mechanism through which ``independent researchers'' could submit proposals to the National Science Foundation, which could in turn compel platforms to disclose data, provided the researcher committed to certain privacy and data security protections \citep{PATA2023}. 
Other initiatives, such as the Social Media Data Act, the Digital Services and Safety Oversight Act, and the Kids Online Safety Act included similar research access proposals \citep{SocialMediaDataAct2021,DigitalServicesAct2022,KidsOnlineSafetyAct2022}.

\paragraph{Summarizing the law of scraping} 
The patchwork of legal requirements that could apply to research involving scraping means that there is no one-size-fits-all analysis, and researchers will need to consider the specific facts of each research proposal to assess their legal risks. 
In general, scraping that involves breaking into online spaces that are not otherwise available to the public will create higher legal risks than scraping only public accessible spaces. 
Privacy considerations will also require researchers to develop plans to appropriately safeguard any personal information they collect and to evaluate and take steps to limit any risks to individuals resulting from the research. 
Additionally, given the importance of data access for sound digital policy-making, legislators -- led by the E.U. -- are starting to propose rules to mandate access to data held by certain platforms for specified public policy purposes.
Taking these together, it is important for researchers to consider the access methods, scope of data collected, and privacy minimization procedures of data stored as they navigate uncertainty around data access as the regulatory environment rapidly shifts. 


\subsection{Ethical Considerations}
The ethics of scraping is not easy to reason about using blanket rules. Within the U.S., the regulatory framework that guides research ethics—45 C.F.R. § 46, known as the ``Common Rule"—provides a robust set of protections for research subjects focused on principles of respect for persons, beneficence, and justice. These principles become operationalized through the formal ethical review of research protocols that must address basic tenets of research ethics, such as protecting the privacy of subjects, obtaining informed consent, maintaining the confidentiality of any data collected, and minimizing harm. 

Yet, with the rise of Internet-based research, traditional ways of assessing the ethics of research falter. Relying on social media data has prompted considerable debate over core issues such as what constitutes ``public" data, whether informed consent is necessary when dealing with ``found data," and even at what stage computational research becomes human subjects research requiring particular ethical protection \citep{metcalf2016human, pater2022no}.
        
One key consideration is consent: the question of whether the subjects of research using scraped data understand they are being studied and have agreed to such data collection and use. Conventional research on human subjects typically involves direct interaction, and thus informed consent can be attained with relative ease. However, just as observing people in public is generally \textit{not} considered human subjects research, institutional review boards typically take the view that studying content publicly available on the internet is \textit{not} considered human subjects research requiring consent. Complicating this further is the lack of consensus among different university institutional review boards on whether and how to review protocols relying on publicly-available online data \citep{vitak2017ethics,zimmer2020ethical}.
        
Meanwhile, organizations such as the Association of Internet Researchers (AoIR) have offered ethics guidelines \citep{franzke2020internet}, and ethics scholars have pushed for researchers engaging with public data to reflect critically on how their data collection might impact users and their communities \citep{zook2017ten,shilton2021excavating} and to be mindful of the original context in which users might be posting data online when considering the appropriateness of scraping data for other uses \citep{zimmer_addressing_2018}. 
        
Central to these recommendations is assessing whether subjects expected that their actions may be observed by researchers: \citet{fiesler2018participant} shows few users of Twitter (now X) are aware that their public tweets are used by researchers. Children are even less likely to fully understand that their data may be used for research, and special precautions should be taken when scraping data from sources that include content created by minors. Research ethics scholars have further pointed to the need to be particularly mindful when collecting online data on sensitive topics or from vulnerable populations \citep{fiesler2016exploring,klassen2022isn}. 
        
Of course, it may be exceptionally difficult to obtain informed consent when collecting platform data, particularly when those data comprise social interactions and social ties between people, or ``relational data.'' For example, social graph and engagement data scraped from social networking sites might implicate secondary users beyond whomever might be ``consenting" to the collection of their data.
Further, images or tags might identify other users who are not able to provide consent.
        
Other ethical considerations when scraping data include understanding whether one's data collection practices place an undue load on websites, affecting performance or causing downtime. An ethical approach to data scraping would generally involve being mindful of the impact on the platform's resources, and using techniques such as rate limiting and respectful crawling behavior to minimize disruption.
        
\subsection{Institutional Considerations}\label{institutional}
As scraping involves significant ethical, legal, and scientific considerations, researchers must navigate diverse challenges and potentially severe risks that likely fall outside of their areas of expertise. In other contexts where such risks appear, universities institutionalize such considerations in order to provide clear guidance, establish uniform processes, and mitigate detrimental outcomes to individuals and the institution. However, given the complexity of issues introduced by scraping, it is often left to the researchers to understand and navigate the various institutional actors that can inform their decision-making. The multi-faceted nature of the institutional support required when scraping, and the fact that the support structures do not typically work together, creates risk for researchers. This section aims to provide guidance for researchers to ensure they are connected to the most important institutional actors.
        
\subsubsection{Institutional Review Boards}
In the U.S., the primary method through which ethical considerations are formalized is the Institutional Review Board (IRB). IRBs were established to protect human research subjects by ensuring that research aligns with federal regulations and guidelines developed to protect human subjects of research, first articulated in the Belmont Report\footnote{See \url{https://www.hhs.gov/ohrp/regulations-and-policy/belmont-report/index.html}} and later codified in U.S. law by way of the Federal Policy for the Protection of Human Subjects, colloquially referred to as the Common Rule (45 CFR Part 46).\footnote{See \url{https://www.hhs.gov/ohrp/regulations-and-policy/regulations/common-rule/index.html}}
        
In the context of reviewing data scraping research protocols, a challenge for IRBs is determining the level of risk associated with scraping digital trace or social media data. If data are publicly available (e.g., posts by a public X account) or have already been collected for other purposes, IRBs generally rule the study either exempt\footnote{See \url{https://www.ecfr.gov/current/title-45/subtitle-A/subchapter-A/part-46}.} or as not engaged in human subjects research. In the ``exempt'' determination, the IRB assesses that the research falls under one of six exempt categories\footnote{Helpfully elaborated on by Michigan State University at \url{https://hrpp.msu.edu/help/required/exempt-categories.html}.} and is thus governed by the ethical principles of the Common Rule, but not by the legal statutes. In the ``not human subjects research'' determination, the IRB determines that the researcher is not obtaining information about living humans through interaction or intervention, or not obtaining identifiable private information and thus falls outside of the Common Rule. 
        
There are two primary ways IRBs differ when evaluating scraping projects. First, IRBs themselves differ between universities, a pattern that is long-standing and well-documented in meta-scientific research \citep{goldman_irb}. To our knowledge there is no systematic study of how different IRBs evaluate scraping, but anecdotal evidence suggests there is wide variation across university IRBs. 

Second, because scraping different websites carries differential risks and benefits, different institutions may focus on different considerations when evaluating a scraping effort for research. For example, scraping public Twitter profiles introduces different risk to user privacy than does scraping private WhatsApp groups; while scraping extremist social media platforms presents different public benefits than scraping GoFundMe pages for medical fundraising. IRBs must assess the risks and benefits of research projects based on, among other factors, the granularity and identifiability of the data, the pre- or post-collection processing methods, the data storage architecture, the data sharing plan for publication or replication, and the scholarly or public importance of the research outputs.\footnote{Again, while the focus of this paper is on U.S. university-based researchers, some international work may be relevant. The Report of the European Digital Media Observatory's working group on Platform-to-Researcher Data Access \citep{EDMO2022} offers GDPR-specific guidance that explicitly focuses on each of these issues for non-scraped data.} 
        
Researchers will benefit from an understanding of IRBs generally\footnote{See, for example \url{https://admindatahandbook.mit.edu/book/latest/index.html}.} and their university's IRB specifically. IRBs should not be seen as a gatekeeper for engaging in ethical research, but rather a partner in ensuring that research aligns with requisite legal, institutional, and ethical standards. 
        
\subsubsection{Legal Review}
The IRB's legal remit is to ensure that research aligns with laws and regulations that govern human subjects research. However, as previously discussed, scraping (potentially) introduces criminal and civil risks under a patchwork of state and federal laws—the relevance of which for academic scraping continues to be adjudicated by the courts. Generally speaking, IRBs do not have the expertise to evaluate and will not consider the legal implications of scraping. Researchers need to understand both the legal risks associated with a particular project and the support the institution might provide if legal action is taken against her. 
These questions are especially exigent given that individual researchers, including students or staff under the supervision of faculty, may face legal risk if they engage in scraping.

A university's Office of General Counsel (OGC) is tasked with providing legal guidance to ensure compliance with the university's legal obligations and inform researcher decision-making. 
Researchers can---but are typically not obligated to---engage with their OGC to evaluate the risks associated with a particular project and get their institution's perspective on those risks. However, it is important that researchers are aware that OGCs exist to represent and protect the university itself. Thus, a researcher undertaking scraping activities that may introduce acute legal risk may want to engage external counsel. Because the ``law of scraping'' remains unsettled—and the level of legal risk varies considerably depending on the precise study design—it will almost never be possible to find a path to ``zero'' legal risk without impeding research aims. Effective counsel can help researchers understand how to minimize those risks, particularly by staying within the bounds of practices that have become commonplace among other actors.

\subsubsection{Technical Review}
While social scientists have increasingly used data scraping and computational methods in research, they have not always kept pace with state-of-the-art methods for collection and privacy-preserving data management \citep{hemphill2021saving}. While data management plans are required by IRBs during the review process, IRBs are generally not responsible nor equipped to provide technical guidance on the state of the art, thus the onus is on the researcher to ensure technical rigor when collecting data via scraping. 
        
Most notably, the ethical and legal risk of any social science research project depends on the technical rigor of data collection, storage, processing, and release. For example, secure data management systems with established access controls and deletion protocols limit the chances of data leakage, which lowers the ethical and legal risks of scraping for research. 
Some universities---especially large, well-funded research institutions---have research IT units that can provide support on key data management questions and provide access to helpful tools and systems. In some cases, universities may provide computing infrastructure that facilitates secure data storage. In contexts where institutional resources are not available, researchers may need to reach out to colleagues in disciplines—such as medicine, public health, and computer science---where technical questions around secure and privacy-preserving data collection have been central to research for some time.

Risks may be reduced further by relying on expert centers of excellence, either within a researcher's own institution, or those available more broadly. For example, many researchers studying digital platforms have hosted data with the Inter-university Consortium for Political and Social Research at the University of Michigan \citep{icpsr-about} or contributed data to the Media and Democracy Data Cooperative \citep{mddc-about}.  

What all this suggests is that researchers should not rely solely on the standard IRB approval process when planning research relying on scraping. Instead, they should work to ensure they have adequate legal protections and that their projects are as technically rigorous as is possible, often through the institutional channels discussed above.

\subsection{Scientific Considerations}\label{scientific}

In addition to the ethical, legal, and institutional considerations we highlight above, researchers must consider scientific issues when scraping data. Many of the issues we highlight in the following section are not unique to scraping as a data collection methodology, but scraping may exacerbate or ameliorate some of the concerns that exist when researchers use APIs or other data collection means.

\subsubsection{Poor sampling frames} 
Perhaps the single largest scientific challenge that researchers face when scraping data is that of their sampling frame. Researchers often have no way of generating random samples based on key units (e.g., users, content, etc.) when they scrape or use an undocumented API, nor do they have the ability to understand the representativeness of the sample they do collect.\footnote{To be clear, this could also be the case with official APIs, but at least with these tools there are often documented descriptions of the API's implementation, which allows data consumers to better understand the biases introduced. Undocumented APIs and scraping typically lack such documentation, which makes sources of bias in the data unknowable.}
For example, a researcher scraping results from an in-product search function (e.g. searching X or Facebook for \texttt{\#covid}) cannot confirm that all users or all content is included. Perhaps more importantly, researchers often have no way of understanding what the sample they've just collected represents compared to the entire corpus. Further complicating the analysis of such scraped data, duplicate content may be difficult to identify, and search results may have passed through moderation filters that are subject to change daily. 

Other technical methods of collection are subject to similarly severe sampling frame issues. Using a browser plugin, for example, may result in significant selection bias---users who understand what a plugin is, how to install it, and what, exactly, their consent is covering are surely not representative of the general public. While these sampling issues can be mitigated to some extent with compensation, thus broadening the populations that are willing to share their data with researchers, it still does not ever allow for researchers to get a representative sample of a given service's users, as only the company knows that information. Thus, even efforts to collect a nationally representative sample may still not have an accurate sampling frame.
	
The key takeaway from these sampling frame issues is that the overall representativeness of research based on scraped data is extremely difficult to confirm. For example \citet{tromble2017we} and \citet{gonzalez2014assessing} find significant variation in Twitter datasets depending on the sampling frame and API endpoints used. With scraped data, however, researchers face a higher level of uncertainty surrounding the sampling frame and data returned, as there is typically no documentation regarding what researchers can expect to reliability collect, as is the case with APIs. Given this limitation, we recommend researchers carefully consider any claims they make with regard to the broad applicability of their findings. 

\subsubsection{Missing data} 
Another scientific issue with scraped data is that some may be missing from collection in a non-random way. Because the data being collected is what is presented to a user, it is likely to be optimized for that purpose instead of completeness \citep{wu2021platform}. Further, platforms may not generate logs accurately, meaning actions the platforms say users took may not be recorded. This is particularly true for data purporting to capture ``views,'' which are often sampled in ways that may result in bias to minimize cost or are optimized for other purposes \citep{TikTok2023VideoPlay,NYT2016VideoCount}.

In addition, the researcher may run into timeouts and/or rate limits when attempting to collect data from an API, which may result in missing entries. Generally speaking, APIs do not respond to queries by delivering data in random order---rather they tend to deliver data in the way most convenient for the developer, which is often the order in which the cases were created. This creates a strong correlation with time, meaning collections with gaps due to rate limiting can miss data concentrated around particular times, further imperiling the representativeness of the resulting data.

\subsubsection{Temporal Instability of Platform Data}

Platforms may update timeline algorithms, follower recommendation systems, user interfaces, features, affordances, and more without any announcement to the users or observers of the platform. These changes can inhibit data collection by breaking pipelines constructed to collect data in a particular format that no longer exists.
They can also inhibit data collection by introducing unobserved and unexplained variation in the data collection process. This contributes to overall challenges with the temporal validity of datasets created or findings drawn from those datasets \citep{munger2023temporal}.

When researchers scrape data, they must contend with the fact that the platform may change at any time. While this is not unique to scraping, the potential for platforms to change in ways that halts data collection altogether is higher with scraping than with APIs. With data collection from APIs, researchers (and other data consumers) can collect data in a predictable, structured way, and the format of the data returned generally does not change without significant notice. However, when platforms change their web interfaces or underlying APIs, it is generally without notice, and can halt data collection that relies on parsing specific fields.

\subsubsection{Construct Validity of Data}
Often API-data collection will result in a mismatch between core concepts and operationalized measurement. Consider using impressions as a measure of media consumption. While they indicate that content was loaded on a device that belongs to the end user, they do not actually measure whether it appeared on a user's screen, much less whether the user actually read it. 

This issue becomes even more complicated when we attempt to measure what the ``modal'' user experience may be on a given website. Many APIs, including some from Google Search and YouTube, deliver results without personalization, which differ from what ``logged in'' users actually see. Results using queries supplied by researchers may be different from how users actually search for content or information on platforms, further limiting the construct validity of such results. 
Such cases require the use of browser plug-ins connected with real users to appropriately measure what is happening to users on-platform, but these methods can run afoul of platform Terms of Service, and in some cases have resulted in legal action by platforms \citep{ortutay2021facebook}.

\section{Recommendations}
Different data access methods necessitate different legal, scientific, ethical, and institutional considerations. Importantly, the same data collected in different ways may invoke different implications. In the following section, we make recommendations for a set of best practices that U.S.-based researchers can follow (and adapt to their contexts). In addition to immediate recommendations for how to conduct research using scraped data \emph{right now}, we also provide recommendations for researchers to improve how we take on data scraping projects in the long term. 

\subsection{Legal}

As legal norms surrounding scraping continue to evolve—and will continue to be tested and refined through litigation targeting the scraping practices of large language model developers (see, e.g., \citet{Krietzberg2024})
—the path forward for researchers will require case-by-case analysis. 
Factors such as the method of data collection, the terms of applicable user contracts, the nature of the services and data being scraped, and the affected jurisdictions will all bear on the legal risks that researchers face when scraping for research purposes.

Nonetheless, legislative developments and recent cases evaluating scraping offer the following guiding principles for researchers looking to gather data via scraping in the near term.  
\begin{itemize}
    \item Websites and services that are accessible to the general public without requiring authentication entail significantly lower legal risk. 
    \item Researchers should carefully consider the potential harms that scraping may pose for individual users and the platforms in question, as material damages can increase the likelihood and value of any claim against the researcher.\footnote{Where scraping does not result in any harm, even if courts find a breach of user contract, a platform's remedies often are limited to blocking future scraping rather than monetary penalties.}
    \item Researchers should collect the minimum data necessary for conducting research and have a credible and documented process for securing the data and protecting privacy. Researchers can analyze the privacy impact of their proposed research through a Data Protection Impact Assessment (see Section~\ref{sect:legal}), which they can use to identify appropriate safeguards for protecting privacy and security.
    \item Finally, researchers focused on ``very large online platforms'' and ``very large online search engines'' should monitor data access developments under the E.U.'s DSA. Tools and mechanisms that platforms and search engines make available to researchers to address E.U. requirements may allow researchers in the U.S. to gain more streamlined access to certain types of data, particularly for research addressing ``systemic risks'' as defined under the DSA.
\end{itemize}

In this evolving landscape of regulation, researchers, IRBs and OGCs will need to get comfortable with some degree of legal uncertainty for the foreseeable future. The goal of legal analysis should be identifying strategies to \textit{manage} legal risks rather than eliminating legal risk entirely. AI companies and other data-driven businesses have developed business models that use scraped data as input, in spite of lingering uncertainties. Excess caution in the face of evolving and uncertain legal norms will lead research data collection practices to fall behind emerging commercial norms, which will slow or block research activities that have the potential to confer wider benefits to the public at large.

\subsection{Ethical}
The rapid growth of projects that rely on scraping large datasets for social and online platforms is testing the ethical frameworks and processes used by researchers and ethical review boards to ensure the protection of human subjects from harm. Our key recommendation for researchers navigating the ethics of scraping data is to ``remember the human'' \citep{fiesler2024remember} and engage in critical ``reflexivity'' \citep{shilton2021excavating} regarding any possible power dynamics that might emerge between the researcher and the users/communities whose data is being scraped. Further, researchers must take steps to acknowledge that the ``publicness'' of data is not the only factor to consider regarding the ethical dimensions of scraping. Rather, we urge researchers to again take a more reflective stance and assess the contextual appropriateness \citep{zimmer_addressing_2018} of collecting, analyzing, and publishing data scraped from a particular platform. This is particularly important when data might be related to sensitive topics or collected from vulnerable communities. 

Along with consulting any available institutional review board or local ethical expertise, numerous resources are available to guide researchers on the ethical complications of scraping data \citep{buchanan_internet_2016}, including guidelines published by the Association of Internet Researchers \citep{markham_ethical_2012,franzke2020internet} and the Pervasive Data Ethics for Computational Research (PERVADE) Data Ethics Tool \citep{pervade_tool}.

\subsection{Institutional}
As we highlight in Section \ref{institutional}, the institutionalization of the ethical, legal, and technical considerations aims to limit risks for researchers, in part by leveraging the expertise that exists across institutional contexts. 
A key challenge for researchers is determining who to engage, when, and towards what ends. 
The Institutional Review Board will ultimately be responsible for ensuring that a project conforms with internal policies and legal standards for research. 
Put simply, we recommend researchers engage their IRB \textit{early and often}.
An IRB should be able to provide guidance on how it understands the risks associated with scraping, as well as connect researchers to other resources from across the university that can help mitigate those risks. 
Those resources include IT services, libraries, and general counsel. 
For researchers engaging in scraping for the first time—or for the first time at a particular institution—colleagues can also act as a resource for understanding how scraping is viewed within that particular environment and how to best navigate the specific actors who are involved in institutionalizing the considerations discussed above. 
    
While a researcher's IRB is ultimately responsible for assessing the risks associated with scraping, researchers might benefit from engaging external resources to navigate the various risks associated with scraping. 
More specifically, IRBs are designed to consider the specific legal and ethical considerations associated with human subjects research, which (as discussed in Section \ref{institutional}) has limitations withing the context of scraping public and semi-public digital trace data. 
However, they may not have expertise in considering the legal and technical dynamics associated with scraping, and not every institution will have the resources for navigating these considerations. 
In these contexts, researchers could consider engaging external groups. 
While these resources will be specific to the particular project, researchers could consider engaging:
\begin{itemize}
    \item Archival organizations, such as the Social Media Archive, which may be able to provide technical advice and infrastructure for projects that involve scraped data.\footnote{For more information, see \url{https://socialmediaarchive.org/?ln=en}}
    \item Professional or disciplinary organizations, which may be able to provide resources with best practices for scraping.
    \item External legal organizations, such as the Knight First Amendment Institute, which may be able to provide legal advice and counsel for projects that involve scraping \footnote{For more information, see \url{https://knightcolumbia.org/}}
\end{itemize}

\subsection{Scientific}
Data access methods have broad implications for the scientific validity and utility of data collected. 
As we highlight in Section~\ref{scientific}, researchers must contend with concerns regarding data availability, sampling frames, missing data, temporal instability of data, and construct validity of the data. 
As a result, extraordinary care must be taken in order to ensure the validity of scientific findings that result from scraped data.  
Three recommendations stand out.

First, we recommend researchers make as much effort as possible for the project to develop a sampling strategy---as opposed to simply using convenience as their guide. 
One laudable example is the way in which The Markup developed a sample for their scraping project, Citizen Browser.  
Unlike many other studies which relied on snowball or other convenience samples for data collection, The Markup \textit{started} their study by recruiting a panel of Facebook users based on a survey vendor's national sample.  
Only then did they install scraping tools connected to those users accounts, which allowed them to appropriately re-weight their data to allow for population inference \citep{citizenbrowser}. 
Because of the rigorous sampling approach The Markup took, they were able to caveat their findings based on biases in their sample. 
Other research projects (e.g. \citet{siegel2021trumping,mcgrady2023dialing}) designed identifier-based sampling strategies to create random (to their knowledge) samples of objects on Twitter and YouTube, respectively.

Second, we recommend researchers carefully caveat claims made based on scraped data. 
There are many factors determining what shows up on a web page or app, particularly when personalization and recommendation systems are in play (e.g., social media feeds). This can mean certain content is over- or under-represented, and that content type may be correlated with unobserved variables (for example, a user's hobbies and interests). This can introduce a great deal of error in any analysis. At the same time, the content that's most widely consumed may not be the content that is easily scraped, and when consumption metrics are available they may not be accurate \citep{TikTok2023VideoPlay,NYT2016VideoCount}.
Often, it is not possible to determine what is missing and why, so it may be challenging to statistically de-bias. 

Third, we recommend researchers think carefully about how research based on scraped data generalizes. 
Because researchers often do not have insights into the underlying systems that they are scraping from, it is not often knowable how much the scraped dataset represents (or does not represent) the underlying data, making claims of generalizibility challenging to verify.
For example, if a researcher were to scrape search results from Twitter related to a specific set of keywords, there would be no way to verify the completeness of the results or the biases introduced by said incompleteness.
Its findings also would not necessarily generalize to other platforms where the audience, affordances, and features may vary. 
These biases in samples can be particularly difficult to correct using de-biasing approaches since most public information about platform users is highly aggregated.

Rather than tossing out any research where researchers are unable to get a representative sampling frame, we recommend researchers carefully caveat their claims based on the attributes of the dataset they are using. 
In the other example, where researchers used identifier-based methods to create random samples of users and/or posts, researchers could caveat that their findings are based on an identifier-based sample, and there may be unknown variation in how platforms assign identifiers.
Researchers could also consider developing well-defined subsamples (e.g. a group of important political actors, celebrities, news networks, or other groups of interest) and collect full data. 
This path would typically yield a smaller dataset than a random sample of all posts or all posts with a given set of keywords, but a well-defined sampling frame could reduce uncertainty surrounding unknown biases in the dataset and could offer a clear basis for researchers to make claims about generalizibility (or lack thereof). 
Because of the limitations of scraping, research based on scraped data should be seen as the first step on a longer journey toward understanding online behavior, with perhaps larger and better samples and/or additional data sources. 

While actionable steps will be specific to the particular research project, we recommend researchers consider:
\begin{itemize}
    \item Collaborating with researchers with experience scraping data or using scraped data in their research 
    \item Careful consideration of whether the question at hand \textit{ought} to be answered using scraping or via some other approach    
    \item An assessment of the literature to ensure the sample needed to answer the right set of questions, especially in the context of scraping data from platforms in which recommendation systems govern content discovery and distribution
    \item Foregoing convenience sampling or snowball sampling in favor of clearly defining a limited sub-population that the data will describe up front (and sampling from it)
    \item Conducting a variety of tests to understand and refine the data pipeline, identify and correct errors and sources of missing data, and deeply understand the representativeness of the dataset being used for research
\end{itemize}

\section{A Researchers' Checklist}
The primary motivation for this paper is to assist researchers in conducting public interest internet research. 
To that end, while we cannot provide a prescriptive set of recommendations that removes all legal, ethical, institutional, and scientific concerns, we can provide a clear list of considerations for which we recommend that researchers have written responses prior to embarking upon a scraping project. 
Further, we believe that documenting a researcher's responses to each of the considerations below will be useful throughout the research process, particularly when engaging with institutional actors as it will help all stakeholders align on the facts and practices of the specific scraping instance. 
Finally, we recommend that researchers use the information generated in this checklist to generate (with the support of their university, if available) a ``communications plan" for how to publicly discuss their activity, particularly in cases where they face backlash for scraping. 
We present this checklist for researchers in the supplementary materials. 

\section{Discussion}
In the longer term, researchers can and should advocate for protections for researchers using scraped data for public interest research when such research is ethically and scientifically rigorous. 
For researchers, this can include communicating with policymakers and the public regarding the importance of research on the Internet, as well as demonstrating high ethical rigor to encourage trust in scientists \citep{dommett2022advocating}. 
Additionally, researchers can engage with coalitions such as the Coalition for Independent Technology Research\footnote{See \url{https://independenttechresearch.org/}} to advocate for policies that protect researchers. 
We recommend researchers look for opportunities to engage across the research community to formalize data security and privacy standards for research involving personal information scraped or otherwise obtained from online sources. 
Data privacy and security standards may ultimately provide a foundation for future data access rights.

To ensure the ethics of research based on scraped data, steps must be taken to fully address the conceptual gaps that persist when research that relies on scraped data often falls outside formal ethical review structures and the educational gaps that might prevent researchers from fully understanding the ethical complexities of scraped data.
One path forward is the continued updates to regulatory policy that governs the protection of human subjects in research activities—such as the Secretary’s Advisory Committee to the Office for Human Research Protections (SACHRP) 2013 publication of new recommendations to the Office of Human Research Protections (OHRP)—to consider emerging ethical concerns with Internet-based research.\footnote{Available at \url{https://www.hhs.gov/ohrp/sachrp-committee/recommendations/2013-may-20-letter-attachment-b/index.html}}  
This body should consider the ethical aspects of the increased use of scraped data and provide updated recommendations.

Academic disciplines and publishers have been increasingly engaging in more direct consideration of ethical dimensions of their methodologies. 
For example, in 2020 the NeurIPS (Neural Information Processing Systems) conference required that all submissions include a statement addressing broader societal impacts or potential harmful consequences of the research, and instituted an ethical review process for papers flagged by reviewers who saw potential ethical concerns. 
More research bodies could embrace these forms of self-governance and establish community norms for the ethical use of scraped data. 
Ethical gaps can be further addressed with longer-term development of enhanced educational resources for both IRBs and researchers alike, incorporating the ethics of scraped data within Responsible and Ethical Conduct of Research (RECR) programs and related research methodology curricula. 

Institutional support for research utilizing scraping could focus on three areas. First, legal questions, both in the U.S. and internationally, will remain a primary consideration for researchers who engage in scraping. 
A key piece is ensuring that institutional stakeholders, such as IRBs and OGCs, understand the legal dynamics in order to provide appropriate guidance and manage individual and institutional risk. 
There are, however, limits to the legal resources provided directly by a researcher's institution—namely, the ability to provide individual legal counsel. 
To that end, legal organizations could consider providing resources directly to researchers engaging in scraping. 
One such model is the Knight First Institute's initiative to support researchers who study social media platforms.\footnote{For more information, see \url{https://knightcolumbia.org/content/knight-institute-to-provide-new-legal-support-to-researchers-studying-online-platforms}}
While this initiative is limited in both topic (platforms) and geography (U.S.), it can serve as a model for similar efforts. Second, disciplinary organizations or university departments could consider offering additional training for researchers who are engaging (or considering engaging) in scraping. 
Instruction should expand beyond a ``how to'' for scraping data, including best practices for data security, risk mitigation, legal and ethical considerations, and more. 
Such efforts could also align with the requirements of academic journals which, anecdotally, have developed diverse (and divergent) assessments of the risks associated with scraped data.
Finally, while researcher projects will inevitably be accompanied by specific considerations, there are general considerations related to scraping. 
Centralizing these considerations would not only increase security and decrease risks, but would also create significant efficiencies for researchers. 
Archives, such as the Social Media Archive, could serve such a role in the ecosystem, and funders—both federal and private foundations—could support such efforts that would support a broad range of grantees.  

Beyond the scope of any one paper, as the field of internet-based research continues to expand and scraping becomes a more commonly-used practice to collect data, it is increasingly important to establish standardized practices around the creation of data through scraping. 
Scientific research benefits from methodologies that ensure data integrity and representativeness, similar to the research practices developed to ensure the reliability of polling. 
Establishing a formal set of guidelines for scraping can enhance the quality and reliability of research data. 
This could be achieved through research dedicated to the development best practices for scraping.
Creating empirically-driven best practices for scraping could ensure that findings based on scraped data meet rigorous scientific standards.
In addition to setting standards and best practices for data collection, researchers should develop specialized training and courses that cater to the evolving needs of internet research, particularly within disciplines where many researchers conduct research on the internet. 
As this area grows, educational institutions play a large role in equipping future researchers with the necessary skills and knowledge. 
This can include not only training in scraping methods, but also in data analysis, ethics, and the application of scraped data in various fields of study.

\section{Conclusions}
In this article, we present a comprehensive framework designed to guide researchers through the complex landscape of legal, ethical, institutional, and scientific considerations essential for scraping projects. 
We examine pertinent case law and U.S. codes, offering an overview the current legal environment.
Additionally, we explore the ethical dimensions of internet research, directing readers to key resources that encourage critical reflection on their proposed studies.
We further discuss the institutional support available to researchers, highlighting both internal mechanisms—such as Institutional Review Boards (IRBs), Research IT teams, and Office of General Counsels (OGCs)—and external experts like the Social Media Archive at ICPSR and the Knight First Amendment Institute. Moreover, we address scientific concerns related to the use of scraped data, emphasizing the importance of assessing external validity and the generalizability of datasets, as well as the limitations of the conclusions that can be drawn.
Overall, this paper equips researchers and institutional stakeholders with the necessary tools and knowledge to conduct scientifically robust research via scraping, ensuring the pertinent legal, ethical, institutional, and scientific factors are thoroughly considered.

\clearpage
\bibliography{references}

\pagebreak
\appendix
\input{appendix}

\end{document}

%% file: appendix.tex
\section{Access Methods}
\label{SI:accessmethods}
While data collection can take an almost infinite number of technical forms, we create four categories for the purposes of this paper.  
As explained below in more detail, the creation of these categories is based primarily on the legal implications of different data collection mechanisms.  
We prioritize legal considerations as the mechanism to cleave data collection methods from one another because we believe they will drive most of the decision-making for researchers.  
However, categorizing these methods also serves to discretize their ethical, scientific, and institutional considerations as well.
We discuss official APIs, which are publicly documented interfaces provided by organizations for data access under specific terms of service. 
Next, we examine undocumented APIs and web scraping, which often lacks formal documentation and may involve different risks and ethical concerns. 
Finally, we cover browser plugins used by enlisted users who voluntarily share their data. 
We also make note of the technical constraints that are relevant to researchers considering collecting data through unsanctioned means. 

    
\subsection{Official APIs}
Application Programming Interfaces (APIs) are mechanisms for moving data from one software system to another. 
Official APIs are APIs that are created and (sometimes) made public by the platforms for sanctioned use by external actors. 
Historically, these systems have not been developed for research, but instead for software developers to extend or embed the functionality of the system offering the API into another application.  
However, researchers have long used these APIs as they offer valuable data that can be more accurate, better documented, and better structured than other forms of data collection.  
Further, because these APIs are developed and maintained by the platform, they are typically well-documented and stable for longer periods of time \citep{parnin2011measuring}.
We define official APIs to make them distinct from the data collection strategies that follow, which are not sanctioned by the platforms themselves. 
        
\subsection{Undocumented APIs}
Undocumented APIs are similar to APIs in that they provide structured data to the user; however, they are undocumented or unofficial, meaning that the public or third-party developers are not the intended users. 
They are typically used by a site or service owner to serve content directly to users via their web page or mobile application, but they can be accessed by anyone who knows the URL and the required parameters.
These APIs can be uncovered by examining the network traffic while navigating a web page and identifying the URLs and parameters that are being used to fetch data and display them to the user \citep{inspect2023browser}. 
Once a researcher understands the API endpoints, query parameters, and data format of an undocumented API, parsing the highly structured data returned (typically JSON or XML data) is relatively simple. 
These APIs can be beneficial in lieu of web scraping because they return structured data and are often more scalable and stable; however, they also have significant legal and scientific downsides for use by researchers. 
Using an undocumented API may violate the terms of service of the website, which could lead to access being blocked or legal consequences (which we discuss further below).
Scientifically, undocumented APIs may return data more reliably, but researchers should use caution when interpreting what the data means or how representative it is, as there is no documentation to confirm that a particular field reflects the information a researcher thinks it does (e.g. a field called ``pubdate'' may be interpreted by the researcher as being the date a piece of content was published, but ``pubdate'' refers to a different value on the website's internal servers).

\subsection{Web Scraping}
Web scraping can be thought of as the inverse of API-based data access: it involves the \textit{creation of structure} from inconsistently or loosely structured data from a website, often by parsing HTML.  
In contrast to scraping via an undocumented API, web scraping is typically more technically complex. 
Web scraping, unlike undocumented APIs, usually returns raw HTML which can be more challenging to parse since the data is less structured.
Websites often change their layout and structure, which can break web scrapers that rely on certain fields being present or in a particular location. 
Central to the idea of web scraping (as opposed to browser-plugin data collection, discussed below) is that the actual data collection is done by a researcher via tools they \textit{own and control}.  
Typically this means a researcher runs software on their own servers, which make requests of a website and parse the resulting HTML (rather than with browser plugins, which are installed by a user and run locally on the user's machines, and information from the browser plugin is relayed to the researcher).  
Web scraping can be done via ``automated'' web browsers---those controllable without using the graphical user interface---that actually render HTML and run JavaScript, or by other tools which collect raw data over HTTP/HTTPS. 
Both web scraping and documented APIs are subject to challenges with logged in and logged out scraping, anti-scraping measures by websites, and issues with rendering static or dynamic pages, which we discuss further below.

\subsection{Browser Plugins}
The final method of data collection assessed in this paper is the browser plugin.  
Like web scraping, browser plugins are tools that allow researchers to capture unstructured data, transform it into that which can be used to answer research questions, and return it to a central repository for analysis.  
Unlike web scraping, though, plugins are not directly deployed by researchers.  
Instead, they are installed by end users of a service into their own web browsers and merely distributed by researchers.  
Put more technically, they're modifying user agents and thus enlisting users in the research data collection process, not unlike citizen science \citep{lazer2020computational}.

Browser plugins result in different trade-offs than web scraping.  
Primarily, because they're operated by end users, they allow for the capture of any data to which that specific user has access (e.g. home timeline feeds on a social media site).  
Further, they are significantly more resistant to anti-scraping techniques, as they appear to services as normal user traffic.  
However, as will be discussed in more detail below, browser plugin data collection can result in highly un-representative samples, violation of data minimization procedures, and legal consequences for the users who install the plugin.

Note that it is critical for researchers to consider the access method of what they're collecting, as these have legal, ethical, and scientific implications for the researchers and any end users they engage in the data collection process.

\section{Technical constraints with unofficial access methods}
\label{SI:technicalconstraints}
Each of the access methods we discuss above have technical constraints, which are implemented by websites to ensure the reliability of their services or protect consumer data. 
When researchers collect data through unofficial access methods, they may encounter challenges with logged in scraping and anti-scraping techniques by websites.

\paragraph{Logged in versus logged out scraping.}
When scraping, the most significant distinction is between scraping content that is accessible to any visitor as a ``logged out'' user, compared with scraping content available only to users who are ``logged in'' to a service. 
When ``logged out,'' scrapers are collecting information that anyone could see and are not identifying themselves to the system they're scraping.  
``Logged in'' scraping, however, often changes the scope and sensitivity of data available.  
On social media websites, for example, ``logged in'' data can include anything an individual user can see---including information only intended for them.  
For example, on X (the social network formerly known as Twitter), an account's followers are not accessible unless one is logged in. 
On some platforms such as Facebook, much of the data of interest to social scientists---for example, posts, ads, and friend data from individuals---is only available when logged in as a user. 
Similarly, the full content of many newspaper articles and other publications may only be available to logged-in users with subscriptions. 
        
This distinction is critically important because it impacts many of the considerations we discuss below, not least of which is legality. 
Not only does logging in provide researchers access to different data---which may change the scientific, ethical, or institutional considerations---but creating accounts with online services generally involves agreeing to a contract (sometimes known as Terms of Service).  
We discuss all these implications below, with special attention to the legal implications in Section~\ref{sect:legal}. 

\paragraph{Static versus dynamic web pages.} There are two types of web pages for researchers to consider: static and dynamic.  Static web pages are those that appear the same for every user, and the appearance does not change while browsing unless the user clicks to a new web page (e.g. Wikipedia is mostly static, as the bulk of content is loaded when a user initially opens the page, and the server does not send requests for more content as the user scrolls). Alternatively, a dynamic web page is typically customized for the user and/or which dynamically loads new throughout the session. Most social media feeds are dynamic web pages; they load HTML for the page skeleton, then content is dynamically loaded and updated as a user scrolls the web page. Essentially, whenever a user gets close to the bottom of the page, another request is made, loading more content for the user to see.  For social scientists, personalized pages are likely the most salient type of dynamic page.  These pages generally require an account and are designed to display specific information for that user. That said, other dynamic pages exist, including those that change based on the IP address of the requester, the requester's imputed location, the time of day, the weather, or indeed almost any other variable designed by the site creator.
Whether a page is static or dynamic impacts the representativeness of the data that the researcher may collect, and collecting data from dynamic pages may require logging in, which may invoke legal implications as well. 

\paragraph{Anti-scraping techniques.} There are numerous technical challenges to collecting data via scraping at scale.  While some are directly designed to stop scraping that site owners do not authorize, others are designed to maintain a quality user experience and in so doing disrupt scraping. 
Often, the primary technical challenge of scraping a given site are the protections sites put in place to prevent distributed denial of service (DDOS) attacks. 
DDOS attacks work by flooding a website with huge numbers of simultaneous requests in order to overwhelm the servers that host it, and thus deny legitimate requests. 
Websites implement protections that often include limiting the number of requests or restricting the amount of data that any computer/IP address can make in a given time frame (typically per minute or per second). 
These protections often result in the rate-limiting of data collection, despite the development of methods to evade them (e.g. using proxies so the IP address looks different for every request). 
For smaller data such as criminal arrest records, where records are typically in the tens of thousands or low millions, it may take several days to collect data from a single IP address. 
For larger data such as social media platform data or campaign finance data, collection via scraping from a single IP address can take months, years, or may simply be impossible to complete. Further, the methods a researcher might use to evade these techniques can have deleterious effects on research, including replicability issues where results are personalized by information from the requesting IP address.  
Further, unlike public APIs, where data providers typically give months or years of notice when making changes to data formats, services often make changes to web page formats explicitly designed to break scraping (e.g. X/Twitter recently made viewing a public profile impossible when a user is not logged in directly to inhibit scraping on the platform). 
Because these changes are rarely announced, they can also result in under-collection of the desired data or failure to collect the desired data. 
Moreover, if a researcher is knowingly evading anti-scraping measures by websites, there may be additional legal consequences. 
Additionally, if researchers are scraping to the extent that they directly impact website performance and the experiences of other users, there are ethical implications as well. 

\section{U.S. Privacy Laws}
\label{SI:usprivacylaws}
The U.S. has no single federal privacy law that applies comprehensive privacy protections across all sectors. 
Rather, a patchwork of federal and state requirements govern the collection and processing of personal information. At the federal level, sectoral laws apply to specific sectors. For example, the Health Insurance Portability and Accountability Act (``HIPAA'') applies to certain ``covered entities'' in the healthcare sector; the Gramm-Leach-Bliley Act (``GLBA'') applies to ``financial institutions''; and the Family Educational Rights and Privacy Act (``FERPA'') applies to ``educational records.''

Outside of these (and other) regulated sectors, the U.S. Federal Trade Commission (``FTC'') has general authority under Section 5 of the FTC Act to police ``unfair and deceptive acts or practices.'' While the FTC has used this authority to shape norms affecting how companies collect, use and share personal information—including in relation to data brokers that accumulate personal information via scraping and other sources \citep{FTC2024}
—the FTC’s authority is limited to acts or practices that are ``in, or affecting, commerce'' \citep{USC15_45}. 
The FTC has not invoked this authority to regulate academic research \citep{Levine2021}.

At the state level, a flurry of lawmaking in recent years, beginning with the California Consumer Privacy Act (``CCPA''), has resulted in a rapid shift in the landscape of privacy protections in the U.S. As of the date of writing, seventeen states have passed consumer privacy laws modeled to varying degrees on the CCPA. These laws generally apply only to entities meeting certain thresholds. For example, the CCPA applies only to ``businesses'' that do business in California and either: (a) have more than \$25 million in annual revenue, (b) buy or sell personal information relating to more than 100,000 California residents, or (c) derive more than 50 percent of revenue from selling personal information \citep{CalCivCode1798140d}.

Most, but not all, of these state privacy laws include important exemptions for non-profit entities. Consequently, academic institutions may be subject to regulation in several states if they meet the relevant thresholds. For example, Colorado's privacy law applies to nonprofits that conduct business in the state, or deliver commercial products or services targeted to state residents, and process personal information relating to at least 100,000 Colorado residents. Privacy laws in Delaware and Oregon—which are not yet in force at the time of writing—exempt only nonprofits with specific missions, such as addressing insurance crime, providing services to victims of or witnesses to certain crimes or violence, or providing programming to radio or television networks. 

While some of these laws could apply to academic institutions in some instances, other exemptions in these laws limit their application to data scraping for research purposes. First, state privacy laws include significant exemptions for ``publicly available information.'' The definition of what qualifies as ``publicly available'' varies, but typically includes any information that a ``consumer has lawfully made available to the general public'' (See, e.g., \citet{ColoradoPrivacyAct130317b}). 
These exemptions will permit the collection of personal information from publicly available websites, but may not exempt data that is available only to a restricted audience pursuant to a user’s privacy settings.\footnote{See, e.g.,  \citet{CalCivCode1798140d}(2) (``For purposes of this paragraph, `publicly available' means: information that is lawfully made available from federal, state, or local government records, or information that a business has a reasonable basis to believe is lawfully made available to the general public by the consumer or from widely distributed media; or information made available by a person to whom the consumer has disclosed the information if the consumer has not restricted the information to a specific audience. `Publicly available' does not mean biometric information collected by a business about a consumer without the consumer’s knowledge.'').}

Second, even if personal information is not publicly available, state laws provide additional exemptions for ``research'' meeting certain conditions. For example, pursuant to regulations issued by the Colorado Attorney-General, the Colorado Privacy Act does not apply to research conducted by a state institution of higher education \citep{ColoradoReg9043Rule202}. 
Similarly, Washington’s My Health My Data Act (``MHMDA''), which regulates ``consumer health data'' and imposes some of the most stringent requirements of any state privacy law, exempts research where it is in the public interest; adheres to all other applicable ethics and privacy laws; is approved, monitored and governed by an institutional review board or other similar entity; and incorporates reasonable (but not specified) safeguards to protect against privacy risks \citep{WaCivCode19373010c}. 

To the extent any such laws do apply to scraping for research purposes—for example, where a researcher is affiliated with a for-profit entity that meets the thresholds of applicability of the relevant laws—researchers generally can comply by maintaining public-facing privacy policies that describe their collection, use, and disclosure of personal information, and by permitting individuals to gain access to a copy of their personal information on request. Researchers may also be subject to other internal-facing requirements regarding how data is collected, analyzed, and stored, such as to implement appropriate security to protect personal information and to apply ``data minimization'' standards by collecting and storing only information that is reasonably necessary for the intended research purposes. These laws also provide individuals with rights to delete personal information, but in many cases, a researcher may have grounds to override the request under various exceptions, provided the research does not involve the disclosure of personal information. 

Additional restrictions may also apply to research involving children’s data or other sensitive data categories, such as data concerning health, and researchers should proceed with caution when scraping facial images and other data that could constitute ``biometric'' information under applicable laws \citep{Sobel2021}.

\section{GDPR}
\label{SI:GDPR}
Individual-level analysis of EU behavior as part of a study could implicate the GDPR, even if the research team has no link to any EU presence. 
The GDPR may apply to researchers where either of two conditions is met: (1) the researcher’s activities are ``inextricably linked'' to an ``establishment'' (i.e., a physical presence) in the EU, such as through the involvement of EU-based researchers or an affiliation with an EU-based institution \citep{EDPB2020} 
; or (2) the research involves ``monitoring the behavior'' of individuals in the EU.\footnote{The GDPR may also apply to entities that ``offer goods or services to individuals'' in the EU, but such activities are unlikely to be relevant to research.} According to regulatory guidance, ``monitoring'' means tracking individuals on the internet, and particularly, conducting ``behavioral analysis or profiling'' based on data collected online \citep{EDPB2020}. 

Key provisions of the GDPR require entities to establish a ``legal basis'' for processing personal information, to provide a privacy notice to individuals whose data the business processes, and to offer those individuals certain privacy rights, such as rights to access, correct, and delete their information, among other internal accountability and documentation requirements. These requirements can pose important challenges for the accumulation of personal information via scraping—and regulators have, in particular, interpreted the GDPR strictly in connection with data brokers that amass large troves of personal information, including from publicly-available sources \citep{Ruschemeier2023}.

In the \textit{research} context, however, the GDPR offers several important exemptions and relaxed requirements that reflect the ``privileged position'' that research occupies within the GDPR. These exemptions may facilitate data scraping for research in many instances \citep{EDPS2020}. 
Specifically:
 
\begin{itemize}
    \item \textit{Researchers can usually rely on the ``legitimate interests'' legal basis under the GDPR to collect and process personal information for research purposes, provided they take steps to protect individuals from privacy risks.}  
\end{itemize}

Researchers can rely on the legitimate interests legal basis where they can demonstrate that processing personal information serves lawful and legitimate purposes without creating undue risks for individuals. While research meeting ethical and methodological standards usually will satisfy this legal basis, researchers bear the burden of documenting, through an internal ``Legitimate Interest Assessment'' that the benefits of processing activities outweigh the risks to rights and freedoms of the individuals whose information the researcher collects \citep{EDPS2020}. 
Regulators have also found that data scraping can meet this ``legitimate interest'' standard, even in the commercial context. 
The European Data Protection Board's task force on ChatGPT concluded that OpenAI could satisfy the legitimate interests test in relation to its web scraping, but that ``adequate safeguards play a special role in reducing undue impact on data subjects and can therefore change the balancing test in favor of the controller'' \citep{EDPBChatGPT2024}.
Appropriate safeguards include ``ensuring that certain data categories are not collected or that certain sources (such as public social media profiles) are excluded from data collection'' and having in place measures ``to delete or anonymise personal data that has been collected via web scraping before the training stage'' \citep{EDPBChatGPT2024}.

Other considerations that affect this analysis will be the extent to which the research zeroes in on and evaluates individual behavior (the more aggregated, the lower the risk level) and the degree to which the research could result in disclosures of private or embarrassing personal information (the GDPR provides incentives for researchers to anonymize or pseudonymize personal information wherever possible to protect against these risks).  Where researchers can demonstrate that the research interest does not pose undue risks for individuals, the GDPR does not require them to seek individuals’ consent \citep{GDPR_Art6}. 

\begin{itemize}
    \item \textit{Researchers do not need to actively notify individuals about their research activities if providing such notice would require disproportionate effort.}  
\end{itemize}
While the GDPR requires ``controllers'' to post a publicly available privacy policy and actively notify individuals of their privacy practices, the GDPR provides exemptions from the active notice requirement where ``the provision of such information proves impossible or would involve a disproportionate effort,'' particularly in the context of research \citep{GDPR_Art145b}.
Information that researchers scrape from websites – where the researcher has no contact with the individual – may qualify for this exception. 
For example, in the EDPB's investigation of ChatGPT, the EDPB concluded that ``it is usually not practicable or possible to inform each data subject about the circumstances. Therefore, the exemption pursuant Article 14(5)(b) GDPR could apply''\citep{EDPBChatGPT2024}. 
Given that regulators have cited research as one of the primary instances where the disproportionate effort exception would apply, researchers that collect personal information via scraping likely can avoid onerous active notice obligations where they have no direct relationship with relevant individuals through which they could provide notice \citep{EDPBTransparency2016}.

\begin{itemize}
    \item \textit{GDPR privacy rights – such as rights to access, correct, delete, object, and restrict – usually do not impair research purposes.}
\end{itemize}

The GDPR generally exempts research from data subject rights to the extent such rights would limit legitimate research activities \citep{EDPS2020}.
For example, the rights to deletion and restriction, which otherwise might allow individuals to opt-out and skew research results, have exceptions that allow researchers to override an individual’s request to preserve the integrity of their research \citep{EDPS2020}. 
It is important to note that, despite these exemptions from certain key requirements, the GDPR does not exempt research entirely. Rather, the law puts the onus on researchers to demonstrate and document how a proposed research design appropriately addresses privacy risks. In particular, to rely on the GDPR’s research exceptions, researchers must implement ``appropriate safeguards [including] technical and organizational measures'' to protect the rights and freedoms of individuals \citep{GDPR_Art891}.
These measures include ``data minimization'' practices to limit the collection of personal information to the bare minimum necessary to achieve the research objective \citep{GDPR_Art891}.
Any research involving personal information should be subject to robust information security practices to protect against unauthorized use and disclosure \citep{GDPR_Art32}.

Researchers will also need to make sure that the personal information they process is reasonably accurate for the purposes of processing and that personal information is retained only for as long as necessary for the intended purpose. 
These requirements are not prescriptive, but researchers must have policies and procedures in place to demonstrate that they have reasonably designed their study to ensure accuracy and to delete any personal information that is no longer needed for the study. 
Researchers should consider and evaluate privacy and data protection risks at the outset before collecting personal information, and they should document how they have managed those risks within a Data Protection Impact Assessment \citep{EDPB2020DPbyDesign}.

Researchers should also exercise additional caution when conducting research that involves collecting sensitive personal information. Specifically, the GDPR applies additional protections to ``special categories'' of personal information, such as data about health, race, religion, ethnicity, religious or political beliefs, and sex life and sexual orientation. To collect and process information about these sensitive attributes—or to infer sensitive attributes from seemingly innocuous personal information—researchers need to satisfy an extra legal condition \citep{noauthor_ot_2022}. 
In the case of scraping for research purposes, the following conditions may support a researcher’s proposed study:

\begin{itemize}
    \item \textit{The GDPR permits researchers to process sensitive data that was ``manifestly made public'' by the relevant individual. }Data scraping may qualify for this exception where the data was made public by the user without restriction, such as on a public platform, where the user has not restricted the audience that can access the relevant content.

    \item \textit{The GDPR also permits researchers to process sensitive data for research purposes where national law provides an exemption.} Most E.U. member states have implemented conditions that permit researchers to process sensitive data for research purposes. For a compilation of member state laws, see the European Digital Media Observatory (``EDMO'') Report of the European Digital Media Observatory’s Working Group on Platform-to-Researcher Data Access \citep{EDMO2022}.
\end{itemize}

If national law does not permit such an exception, researchers may be able to limit their legal risk by immediately scrubbing relevant data of personal identifiers and/or sensitive attributes. For instance, in the context of the EDPB's investigation into OpenAI's practices, the EDPB concluded that OpenAI could protect against the risks of sensitive data ingestion through its scraping by implementing ``safeguards [such as] filtering data categories falling under Article 9(1) GDPR [at] both, data collection (for example, selecting criteria for what data is collected) and immediately after data collection (deleting data)'' \citep{EDPBChatGPT2024}. 

\section{Researcher Checklist}
\noindent\textbf{\underline{Research Purpose}}
\begin{checklist}
    \item Research question, hypothesis, and other relevant background
    \item Statement of broad social value of the research, particularly in light of any potential privacy or other harms that could come to both individuals and groups as a result of the research
\end{checklist}
\noindent\textbf{\underline{Data Collection}}
    \begin{checklist}
        \item Identification of the sources from which the data will originate
    \item Description of the specific methods by which the researcher will collect the data
    \item Identification of any user contracts (e.g., terms of service; end user license agreement; etc.) that apply to the data source, and any actions that will be required to accept such contracts (e.g., logging into a service; checking a box to accept the user contract; etc.) in order to gain access to the data for scraping
    \item Description of the relevant data categories collected for the research purposes and the methods the researcher will employ to limit collection to only relevant categories
    \item Identification of any sensitive data categories that may be collected, including personal information concerning health, race, religion, ethnicity, political beliefs, sexual orientation and sex life, and biometric data 
    \item Description of the ``publicness" of the data to be collected and any privacy settings applicable to the relevant data to restrict access, along with the context in which it was created---and the extent to which that context is preserved or violated by the planned research
    \item Description of intended collection time frame
    \end{checklist}
\noindent\textbf{\underline{Scientific Validity}}
    \begin{checklist}
        \item Description of the intended sampling strategy and the resulting generalizability of any findings based on it
        \item Description of data that is known to be missing, and that which one can reasonably suspect to be missing, given the collection method and sampling strategy
        \item Description of the construct validity of the data being collected, with justifications for that validity
        \item Description of any strategy to manage temporal instability, if required
    \end{checklist}
\noindent\textbf{\underline{Proportionality and Risk Mitigation}}
    \begin{checklist}
        \item Description of the necessity of scraped data for the research purpose, and justification for seeking such data via scraping rather than other means
        \item Identification of potential risks to individuals as a result of scraping and research analysis, including risks relating to profiling, behavioral analysis, data leakage, disclosure of private facts, embarrassment, manipulation, etc.
        \item Analysis of ethical considerations associated with data access method and research design
        \item Description of safeguards employed to protect privacy and manage identified risks, such as data masking and pseudonymization, data minimization, access controls, etc.
        \item Description of applicable data security safeguards to protect against unauthorized access and use, including designated data repositories, limitations on authorized access, auditing and logging of access, and contractual commitments
    \end{checklist}
\noindent\textbf{\underline{Stakeholder and Institutional Engagement}}
    \begin{checklist}
        \item Documentation for a meaningful review by in IRB or equivalent---for which material from this checklist should be provided
        \item Documentation for engaging with independent experts, such as data archives or professional organizations
        \item A plan for what information (e.g. IRB approval or exemption, etc.) should be public before the work begins, and plans for reactive communications responses, if required, once the work begins.
    \end{checklist}